\documentclass[]{rnaastex}

\usepackage{graphicx}
\usepackage[suffix=]{epstopdf}
\usepackage{natbib}
\usepackage{amsmath}
\usepackage{xspace}
\usepackage[overload]{textcase}
\usepackage{hyperref}

\newcommand{\lb}{\lambda_C}
\newcommand{\ld}{\lambda_L}

\begin{document}

\title{A STATIONARY DRAKE EQUATION DISTRIBUTION \\AS A BALANCE OF BIRTH-DEATH PROCESSES}

\correspondingauthor{David Kipping}
\email{dkipping@astro.columbia.edu}

\author[0000-0002-4365-7366]{{\fontsize{10.5}{12.6}\selectfont \textcolor{black}{David Kipping}}}
\affil{Department of Astronomy,
Columbia University,
550 W 120th Street,
New York, NY 10027, USA}

\begin{abstract}
Previous critiques of the Drake Equation have highlighted its deterministic
nature, implying that the number of civilizations is the same at all times.
Here, I build upon earlier work and present a
stochastic formulation.The birth of civilizations within the
galaxy is modeled as following a uniform rate (Poisson) stochastic process, with a mean
rate of $\lb$. Each then experiences a constant hazard rate of collapse, which
defines an exponential distribution with rate parameter $\ld$. Thus, the galaxy
is viewed as a frothing landscape of civilization birth and collapse. Under
these assumptions, I show that $N$ in the Drake Equation must follow another
Poisson distribution, with a mean rate $(\lb/\ld)$. This is used to
demonstrate why the Copernican Principle does not allow one to infer $N$,
as well evaluating the algebraic probability of being alone in the galaxy.
\end{abstract}

\keywords{extraterrestrial intelligence}


\section{A Stochastic Formalism} 

The \citeauthor{drake:1965} Equation formulates the number of communicative
civilizations in the galaxy, $N$, as product of 1) the star formation rate,
2) numerous conditional probabilities concerning life, and 3)
the lifetime, $L$, of said civilizations. More succinctly, it's
the rate at which communicative civilizations emerge multiplied by their
lifetime:

\begin{align}
N = \Gamma_C L.
\end{align}

The formulation wasn't originally intended
as a true calculator \citep{drake:1991}, rather as a pedagogical and
organizational framework. A basic limitation is that for any specific choice
for the inputs, $N$ is a fixed number - implying that at all
times there is precisely the same number of civilizations
\citep{cirkovic:2004}. A better interpretation of the Drake Equation, then,
is that it describes the \textit{mean} number of civilizations, cast as
the mean rate of emergence multiplied by the mean
lifetime i.e. $\mathrm{E}[N] = \mathrm{E}[\Gamma_C] \mathrm{E}[L]$.

Efforts to update the Drake Equation, from its original deterministic form to
a probabilistic one, have been previously proposed \citep{forgan:2011}.
For example, \citet{maccone:2010} highlight that given a lengthy series of
multiplicative distributions, the Central Limit theorem dictates
that $N$ is log-normally distributed - although following our
earlier argument it's more accurate to state that $\mathrm{E}[\Gamma_C]$ is
log-normally distributed. To make $N$ truly stochastic, rather than
$\mathrm{E}[N]$, \citet{glade:2012} (G12) proposed that a homogeneous Poisson
process is suitable to describe the births of new civilizations.

A homogeneous Poisson process describes a stochastic process whose success rate
is a time-independent quantity. If one chooses a recent epoch of
${\lesssim}100$\,Myr, the galaxy can be argued to hardly evolve and thus it's
reasonable to suggest that the mean rate of emerging civilizations is
approximately constant during this time. Although the galaxy
is made up of many different star-types, ages and environments, after averaging
over the ${\sim}10^{11}$ examples, the mean rate of emergence does not evolve. 

In this note, I highlight that the formulation of G12 can be expanded to reveal
a simple yet reasonable stochastic Drake Equation. Specifically, I adopt a
stochastic model for the lifetime of civilizations following
\citet{kipping:2020}, who propose an exponential distribution. The basic
assumption, similar to the Poisson process, is that the existential risk is the
same in any given time interval (e.g. the probability of a fatal gamma ray
burst \citep{piran:2014} doesn't change in time).

Let us denote that the rate of civilization birth, $\Gamma_C$, follows a Poisson
distribution characterized by the mean rate parameter
$\lb(\equiv\mathrm{E}[\Gamma_C])$, such that the mean number of births in a time
$t$ is $\lb t$. Next, let us write that lifetime of civilizations,
$L$, follows an exponential distribution characterized by a rate parameter
$\ld$, such that the mean lifetime of civilizations is $\mathrm{E}[L]=1/\ld$.
The Drake Equation is thus governed solely by $\Gamma_C$ and $L$, where
$\Gamma_C \sim \mathrm{Po}[\lb]$ and $L \sim \mathrm{exp}[\ld]$, but what is
the probability distribution for $N$ then?

Consider an infinitesimal time interval, $\delta t$. Let us write that the mean
number of civilization births in this interval is $\delta n$. Via the
theorem of linearity of expectation, the mean number of extant civilizations
in this interval, $\mathrm{E}[N]$, must equal $\delta n$ \textit{plus} the
mean number of survivors from earlier intervals. In the previous interval, the
mean number of births must also be $\delta n$ (by definition of a Poisson
process), but not all will survive. If one writes the risk of death as
$\delta d$, then the mean number of survivors will be $\delta n (1-\delta d)$
from the previous cycle, $\delta n (1-\delta d)^2$ from the cycle before,
etc. Summing over an infinite number of intervals one finds a convergent series
where

\begin{align}
\mathrm{E}[N] &= \sum_{i=0}^\infty \delta n (1-\delta d)^i = \frac{\delta n}{\delta d}.
\end{align}

Let us now turn to $\delta d$. An exponential distribution is defined by a
constant hazard function, such that the mean probability of failure in a time
interval $\delta t$ equals $\ld \delta t$. Using this, and taking the
infinitesimal limit, one has

\begin{align}
\mathrm{E}[N] &= \frac{1}{\ld} \frac{\partial n}{\partial t}.
\end{align}

Note that $(\partial n/\partial t)$ defines the mean number of
civilizations birthed per unit time, which by definition equals $\lb$, thus:

\begin{align}
\mathrm{E}[N] &= \frac{\lb}{\ld}.
\end{align}

\begin{figure*}
\begin{center}
\includegraphics[width=15.5cm,angle=0,clip=true]{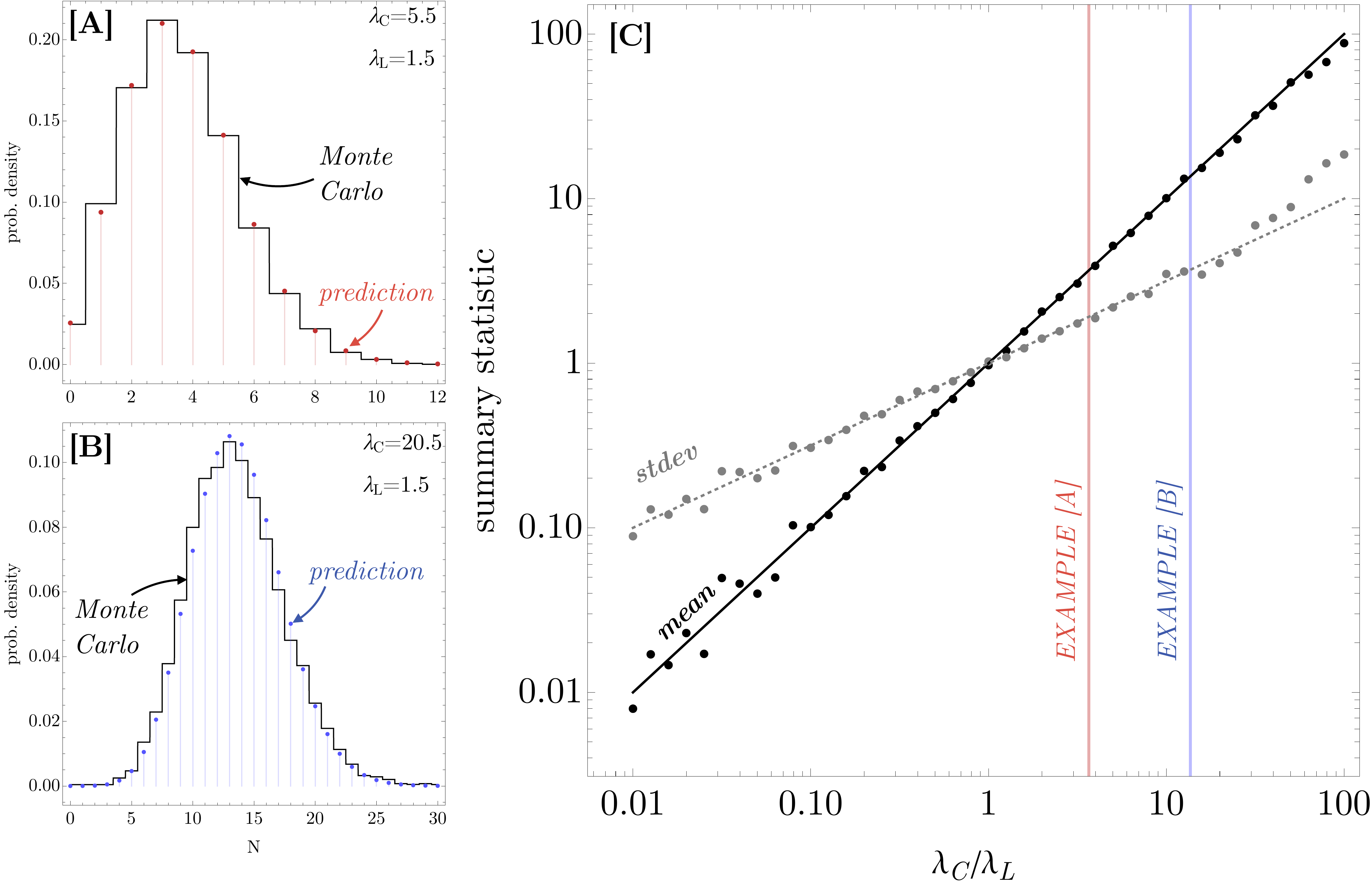}
\caption{\emph{
Panels A and B show a histogram (black) of the number of civilizations present
in each snapshot of a Monte Carlo simulation with $10^6$ time steps with
$\lb=5.5$ \& $\ld=1.5$ for A and $\lb=20.5$ \& $\ld=1.5$ for B. The colored points
show the prediction of our stochastic distribution. Panel C shows the
mean (black) and standard deviation (gray) of a broader set of examples, with
the predictions shown as lines through each.
}}
\label{fig}
\end{center}
\end{figure*}

Since the time interval $\delta t$ is arbitrary, the above is true at all times
and thus defines a Poisson process with a rate parameter
$\lambda_N \equiv \lb/\ld$. Accordingly, one can write our new stochastic
Drake Equation as $N \sim \mathrm{Po}[\lb/\ld](=\mathrm{Po}[\lambda_N])$. This
remarkably simple form\footnote{
It is noted that an alternative derivation is possible exploiting the fact
that for independent random variables the mean of the product equals the 
product of the means.
}
highlights how the number of civilizations present over various time intervals
follows the classic Poisson distribution governed by just two terms - a
balancing act of birth and death. This expression was verified through Monte
Carlo experiments depicted in Figure~\ref{fig}. This formulation presents
several key insights highlighted in what follows.

\section{The Abject Failure of the Copernican Principle}

The Copernican Principle is often used to motivate the plurality of life (e.g.
\citealt{westby:2020}). However, using our formulation, a complete failure of
this principle is revealed, as a result of selection effects. The key is that
any self-aware entity is living in a galaxy with at least one success, it
cannot reside within the realization of $N=0$. Consequently, $N=0$ cases
must be discounted, and so self-aware entities must be drawn from the
\textit{truncated} Poisson distribution

\begin{equation}
\mathrm{Pr}(N|\mathrm{selfaware}) =
\begin{cases}
\frac{1}{\exp(\lambda_R)-1} \frac{\lambda_R^N}{N!} & \text{if } N \geq 1, \\
0  & \text{if } N = 0 .\\
\end{cases}
\label{eqn:truncated}
\end{equation}

The Copernican Principle makes inferences about others based on itself as an
example, and here the inference would be about $\lambda_N$, which controls $N$.
Accordingly, one might now attempt to infer $\lambda_N$ given one's own
existence - such that the ``data'' we are conditioning upon is that $N \geq 1$.
Via Bayes' theorem, one has

\begin{align}
\mathrm{Pr}(\lambda_N|N\geq1,\mathrm{selfaware}) \propto \mathrm{Pr}(N\geq1|\lambda_N,\mathrm{selfaware}) \mathrm{Pr}(\lambda_N|\mathrm{selfaware}),
\end{align}

where the ``selfaware'' conditional is explicit throughout. The
first-term on the right is known as the likelihood, and governs how the data
informs our inference. In this case, the likelihood function can be solved
analytically as

\begin{align}
\mathrm{Pr}(N\geq1|\lambda_N,\mathrm{selfaware}) &= \sum_{N'=1}^{\infty }\mathrm{Pr}(N'|\mathrm{selfaware}).
\end{align}

Using Equation (\ref{eqn:truncated}) and summing over all $N'$ indices
returns 1 - which is no surprise because the distribution is normalized over
the interval $[1,\infty]$ by virtue of its truncation. Thus,
the likelihood function is a constant and contains no information about
$\lambda_N$, and so inferences using the Copernican Principle are meaningless.

\section{Are We Alone?}

Using our stochastic distribution, one can straight-forwardly calculate the
probability that we are not alone in the galaxy. As before, since we know that
we are self-aware, then probabilities using this work's formulation should
account for this and use Equation~(\ref{eqn:truncated}) - the truncated form.
Accordingly, the probability that we are not alone (within some given volume
governed by $\Gamma_C$) is

\begin{align}
\mathrm{Pr}(N\geq2|\mathrm{selfaware}) &= 1 - \frac{\lb/\ld}{e^{\lb/\ld}-1}
\end{align}

which is better than 0.5 for all $(\lb/\ld) > 1.26$ and approaches
unity as $(\lb/\ld) \gg 1$. The $\ld$ rate is broadly unknown but,
as an example, \citet{simpson:2016} suggests $\ld = 0.002$/year using 
the Doomsday Argument. In such a case, we are likely not
alone if $\lb\gg0.2$/century.

Finally, it's highlighted that the mean number of civilizations who've ever
arisen over a time interval $T$ is $\lb T$, whereas the current number is
$\lb/\ld$, and thus if $\lb T \gg (\lb/\ld)$, or simply $T \gg \mathrm{E}[L]$,
then one expects there to be far more extinct civilizations that extant,
in which case one might place greater weight on artifact SETI searches.

\acknowledgments

DK thanks the supporters of the Cool Worlds Lab: Tom Widdowson, Mark Sloan, Douglas Daughaday, Andrew Jones, Jason Allen, Marc Lijoi, Elena West, Tristan Zajonc, Chuck Wolfred, Lasse Skov, Geoff Suter, Max Wallstab, Methven Forbes, Stephen Lee, Zachary Danielson, Vasilen Alexandrov, Chad Souter, Marcus Gillette \& Tina Jeffcoat.



\begin{thebibliography}{}
\expandafter\ifx\csname natexlab\endcsname\relax\def\natexlab#1{#1}\fi

\bibitem[\protect\citeauthoryear{{\'C}irkovi{\'c}}{2004}]{cirkovic:2004}
{\'C}irkovi{\'c}, M., 2004, AsBio, 4, 225

\bibitem[\protect\citeauthoryear{Drake}{1965}]{drake:1965} 
Drake, F. 1965, The Radio Search for Intelligent Extraterrestrial Life, eds. Mamikunian G,
Briggs MH. pp. 323–345.

\bibitem[\protect\citeauthoryear{Drake \& Sobel}{1991}]{drake:1991} 
Drake, F. \& Sobel, D., 1991, Is Anyone Out There?, Simon and Schuster, London.

\bibitem[\protect\citeauthoryear{Forgan}{2011}]{forgan:2011}
Forgan, D.~H., 2011, IJAsB, 10, 341

\bibitem[\protect\citeauthoryear{Glade et al.}{2012}]{glade:2012}
Glade, N., Ballet, P., Bastien, O., 2012, IJAsB, 11, 103

\bibitem[\protect\citeauthoryear{Kipping et al.}{2020}]{kipping:2020}
Kipping, D., Frank, A., Scharf, C., 2020, IJAsB, 19, 430

\bibitem[\protect\citeauthoryear{Maccone}{2010}]{maccone:2010}
Maccone C., 2010, AcAau, 67, 1366

\bibitem[\protect\citeauthoryear{Piran}{2014}]{piran:2014}
Piran, T. \& Jimenez, R., 2014, Phys. Rev. Lett., 113, 231102

\bibitem[\protect\citeauthoryear{Simpson}{2016}]{simpson:2016}
Simpson, F., 2016, arXiv, arXiv:1611.03072

\bibitem[\protect\citeauthoryear{Westby \& Conselice}{2020}]{westby:2020}
Westby, T. \& Conselice, C.., 2020, ApJ, 896, 58

\end{thebibliography}
\end{document}